\documentclass[conference]{IEEEtran}

\usepackage{amsfonts}
\usepackage{algorithmic}
\usepackage{algorithm}
\usepackage{amssymb}
\usepackage{amsthm}
\usepackage{times,color}
\usepackage[cmex10]{amsmath}
\usepackage{cases}
\usepackage{graphicx}
\usepackage{float}
\usepackage{color}
\usepackage{graphicx}
\usepackage{times,color}
\usepackage[draft,ulem=normalem]{changes}
\newtheorem{theorem}{Theorem}

\newtheorem{lemma}{Lemma}

\newtheorem{claim}{Claim}
\newtheorem{example}{Example}

\newcommand{\comment}[1]{}

\newcommand{\cR}{{\cal R}}
\newcommand{\cC}{{\cal C}}

\newcommand*{\underuparrow}[1]{\ensuremath{\underset{\uparrow}{#1}}}
\IEEEoverridecommandlockouts

\begin{document}
\title{Balanced Permutation Codes}
\author{
  \IEEEauthorblockN{
    Ryan~Gabrys\IEEEauthorrefmark{1}\IEEEauthorrefmark{2}~and
    Olgica~Milenkovic\IEEEauthorrefmark{1}}
  {\normalsize
    \begin{tabular}{ccc}
      \IEEEauthorrefmark{1}ECE Department, University of Illinois, Urbana-Champaign~~ &
      \IEEEauthorrefmark{2}Spawar Systems Center, Pacific~~ \\
    \end{tabular}}\vspace{-3ex}
    }


%


\maketitle

\begin{abstract}
Motivated by charge balancing constraints for rank modulation schemes, we introduce the notion of balanced permutations and derive the capacity of balanced permutation codes. We also describe simple interleaving methods for permutation code constructions and show that they approach capacity. \end{abstract}
%
\IEEEpeerreviewmaketitle

\vspace{-0.5ex}\section{Introduction}
We consider a new constraint on permutations that requires moving averages of symbols to be closely concentrated around the mean running average. This constraint and the resulting coding schemes aim to balance charges across cells in rank modulation systems for flash memories~\cite{JMSB09}. Balanced codes may also potentially aid in detecting and correcting errors. This can be accomplished by monitoring whether the balancing constraint is satisfied during the decoding process. Hence, the constraint complements existing constraints imposed so as to contain cross-leakage between neighboring cells~\cite{SY14}\footnote{These constraints often go under the name two-neighbor constraints, and will be discussed in more detail at the end of the paper}. The permutation balancing constraint may also be seen as a block-by-block extension of classical bounded running digital sum codes~\cite{WISS13}, and some of the construction ideas pursued were inspired by the well known Knuth's balancing algorithm~\cite{K86}. 

The constraint may be succinctly described as follows. 

Let $n$ be a fixed positive integer, and let $[n]=\{{1,2,\ldots,n\}}$. For any two non-negative integers $a \leq b$, we use $[a,b]$ to denote the set $\{{a,a+1,\ldots,b-1,b\}}$ and we use $\mathbb{S}_n$ to denote the set of permutations of length $n$.
We are concerned with studying codes $\cC(n,S) \subseteq \mathbb{S}_n$, where $S \subseteq [1,n]$, defined as follows: $\pi \in \cC(n,S)$ if and only if for every $b\in S$ and ${1} \leq j \leq n-b+1$, one has
\vspace{-0.5ex}\begin{align}\label{eq:constraints}
\frac{n+1}{2} (1- D(b,n)) & \leq \frac{\pi(j) + \pi(j+1)+ \cdots +\pi(j+b-1)}{b} \\
& \leq  \frac{n+1}{2}(1 + D(b,n)). \nonumber \vspace{-0.5ex}
\end{align}
Here, $D(b,n)$ takes values in the interval $[0,1]$ and is allowed to be a function of $b$ and $n$. We refer to the constraint in \eqref{eq:constraints} as a \emph{balancing constraint}, as it requires any $b$-consecutive sum of elements in a permutation to stay close to the mean $b$-consecutive sum, which equals $b(n+1)/2$.

A constraint related to the balancing constraint was studied under the name of \emph{low-discrepancy permutation constraint} in a handful of papers~\cite{AFG02}. The authors of~\cite{AFG02} studied the problem of finding \emph{a smallest discrepancy permutation}. In particular, they defined the discrepancy of a permutation $\pi \in \mathbb{S}_n$ according to
\vspace{-0.5ex}\begin{equation}
disc(\pi,b)=\max_{1 \leq i \leq n-b}\; \left|  \sum_{j=1}^{b}\, \pi_{i+j} - b\,\frac{n+1}{2}\right|, \notag \vspace{-0.5ex}
\end{equation}
and focused their attention on computing $disc(n,b)=\min_{\pi \in \mathbb{S}_n} \, disc(\pi,b).$
They showed that  $disc(n,b)\leq 2$ for any choice of the parameters $n$ and $b>1$.
In contrast to the work pertaining to permutation discrepancy, our work is not concerned with finding the smallest discrepancy permutation but rather with \emph{constructing codes} of relatively small discrepancies and of size as large as possible. Furthermore, the discrepancy constraint places a constraint on substrings of fixed length $b$, whereas 
the balancing constraint in~(\ref{eq:constraints}) is more general as balancing is required for all blocks of length $b \in S$, where $S$ may contain more than one value. This code design approach offers certain advantages in terms or built-in error detection capabilities, and in addition, it covers the single blocklength $b$ balancing constraint by definition. To the best of the authors knowledge, this coding problem has not been studied before in the literature.

One use of the results of~\cite{AFG02} allows us to show that for $S=[2,n]$, $\cC(n,S)$ is non-empty provided that $D(b,n) \geq 4/(b(n+1))$.  As an example, it is easy to see that the smallest discrepancy set of permutations for $n=4$ and $b=2$ equals:
\vspace{-0.5ex}\begin{align*} 
&(1,3,2,4),\; (1,4,2,3), \; (2,3,1,4),\; (2,4,1,3),\\
&(3,1,4,2),\; (3,2,4,1), \; (4,1,3,2),\; (4,2,3,1). \vspace{-0.5ex}
\end{align*}
The discrepancy of the permutations is $1$. Allowing for a larger discrepancy clearly increases the size of the permutation code.

For simplicity, our focus will be on two special cases for the set $S$, $S_1=[1,n]$
 and $S_2=\{{2,4,\ldots,2 \cdot s\}}$, for some $s$, such that $s < \frac{n}{\lceil n^{\epsilon} \rceil}$, and their corresponding $D(b,n)$ functions,
\vspace{-0.5ex}\begin{align}\label{eq:c1}
D_1(b,n) =  \frac{4}{b},\;\; \text{and} \;\; D_2(b,n)=\frac{16}{b\,n^{\epsilon}},\vspace{-0.5ex}
\end{align}
where $\epsilon \in (0,1)$. These two constraints were selected to demonstrate that one can balance out the same set of permutations for almost all choices of $b$ simultaneously, although not with the same balancing function $D(b,n)$ - the balancing function is weaker for smaller $b$. For the first balancing constraint where $D(b,n)=D_1(b,n)$ and $S=S_1$, the resulting permutation code has capacity one. On the other hand, it is impossible to have the same permutation code have a discrepancy uniformly scaling with $n^{-\epsilon}$ for all choices of $b \in [1,n]$. Some divisibility properties on $b$ have to be imposed. At the same time, the code rate is strictly less than one, equal to $1-\epsilon$.   

Codes $\cC(n,S)$ with codewords satisfying (\ref{eq:constraints}) and discrepancy $D_1$ and $D_2$ are referred to as a $D_1$- and $D_2$-balanced permutation codes. Let
\vspace{-0.5ex}\begin{align}\label{eq:cap}
\cR_{i}=\lim_{n \to \infty} \sup \; \frac{\log \, |\cC(n,S)|}{ \log\, n! },\vspace{-0.5ex}
\end{align}
for $i=1,2,$ denote the capacity of a $D_i$-balanced permutation code expressed in bits\footnote{Here and throughout the paper we assume that all logarithms are evaluated base two.}. 
Our main results demonstrate that when $S=S_1$, $\cR_1 = 1$.  For $S=S_2$, we get that $\cR_2 = 1-\epsilon$. We also provide a sampling of results for  $D_1-$ and $D_2-$balanced codes that satisfy the two-neighbor constraint introduced in~\cite{SY14}.

\section{The Capacity of $D_1$-Balanced Codes}\label{sec:constructd1}
In this section, we present a construction for $D_1$-balanced permutation codes that achieve rate one. For simplicity of exposition, we suppose that $n$ is an even number. 

The idea behind the code construction is to partition the set $[n]$ into two subsets, $P_1=[n/2]$ and $P_2=[n] \setminus [n/2]$. The symbols in the set $P_1$ are arranged according to a permutation $\gamma_1 \in \mathbb{S}_{\frac{n}{2}}$, and the resulting sequence is denoted by $O_1$. Similarly, the symbols in the set $P_2$ are arranged according to a permutation $\gamma_2 \in \mathbb{S}_{\frac{n}{2}}$, and the resulting sequence is denoted by $O_2$. We form permutations $\pi \in \cC(n,S_1)$, which we subsequently prove to be $D_1$-balanced as follows. 

We initialize the construction by choosing the first element of $\pi$ to be the first element of $O_1$; we consequently remove that element from $O_1$. The second element of $\pi$ is set to the first element of $O_2$ and this element is subsequently removed from $O_2$. Suppose next that $j-1,$ $j>1$, symbols of $\pi$ have been selected. To determine the next symbol in $\pi$, we compute the accumulated average sum $A(j-1)=\frac{1}{j-1} \sum_{\ell=1}^{j-1} \pi(\ell)$. If $A(j-1) < \frac{n+1}{2}$, we set the $j$-th element of $\pi$ to be equal to the first element of $O_2$ and remove the element from $O_2$. If $A(j) \geq \frac{n+1}{2}$, then we set the $j$-th element of $\pi$ to be equal to the first element of $O_1$ and remove this element from $O_1$.

The procedure is illustrated via the example below.

\begin{example} \label{ex:d1encode}
Let $n=12$. In this setting, we have $P_1=\{1,2,3,4,5,6\},$ $P_2=\{7,8,9,10,11,12 \}$. 
By choosing $\gamma_1=(3,4,1,2,5,6)$ and $\gamma_2 = (6,5,4,3,2,1)$, we arrive at $O_1=(3,4,1,2,5,6)$ and $O_2=(12,11,10,9,8,7)$.

For the given choice of $\gamma_1,\gamma_2,$ we initialize $\pi$ as $ \pi = (3,12,\ldots)$
and obtain $O_1=(4,1,2,5,6)$, $O_2=(11,10,9,8,7)$. Next, we evaluate $A(2) = \frac{1}{2} (3 + 12) =7.5 \geq 6.5$, and subsequently select the first element from $O_1$ to obtain $\pi = (3,12, 4,\ldots)$. We then compute $A(3) = 6.33$ and arrive at 
$\pi = (3,12,4,11,\ldots)$. Continuing until all elements are used up, 
we obtain $(3,12,4,11,1,10, 2, 9, 8, 5, 7,6)$.
\end{example}

It is straightforward to see  that if $\pi$ is constructed according to the previous procedure using two permutations $\gamma_1, \gamma_{2}$, while $\pi'$ is constructed from $\gamma_1', \gamma_{2}'$ where $(\gamma_1, \gamma_{2}) \neq (\gamma_1', \gamma_{2}')$, then $\pi \neq \pi'$. Therefore, the cardinality of $\cC(n,S_1) \subseteq \mathbb{S}_n$ equals the number of possible choices for the permutations  $\gamma_1, \gamma_{2},$ i.e., $ |\cC(n,S_1)| = \left( \frac{n}{2} ! \right)^{2},$
which implies that $\cR_1=1$. 

Note that the sequences $O_1, O_2$ are updated after each step, i.e., after each extension of the permutation $\pi$. For notational convenience, we let $O_1^{(j)}$ denote the sequence $O_1$ after $j$ elements have been added to the permutation $\pi$; similarly, we let $O_2^{(j)}$ denote the sequence $O_2$ after $j$ elements have been added to the permutation $\pi$. For ease of notation, we use $len(O)$ to denote the length of the sequence $O$. The next lemma establishes an important property of the encoding procedure described in Example~\ref{ex:d1encode}.

\begin{lemma}\label{lem:correctness} For any $\pi \in \cC(n,S_1)$ constructed using two given permutations $\gamma_1,\gamma_2,$ and any $j \leq n$, $\frac{1}{j-1} \sum_{\ell=1}^{j-1} \pi(\ell) < \frac{n+1}{2},$
implies that $len(O^{(j-1)}_2) \neq 0$. Similarly, 
$\frac{1}{j-1} \sum_{\ell=1}^{j-1} \pi(\ell) \geq \frac{n+1}{2},$
implies that $len(O_1^{(j-1)}) \neq 0$.
\end{lemma}
\begin{IEEEproof} Suppose that
$\frac{1}{j-1} \sum_{\ell=1}^{j-1} \pi(\ell) < \frac{n+1}{2}$
and that on the contrary, $len(O^{(j-1)}_2) = 0$. Let $P$ represent the set of symbols in the sequence $(\pi(1), \ldots, \pi(j-1))$. The set $P$, and the sets of elements contained in $O^{(j-1)}_1$ and $O^{(j-1)}_2$ form a partition of $[n]$. Clearly, the average symbol value of the set $[n]$ equals $\frac{n+1}{2}$. Hence, if $len(O^{(j-1)}_2) = 0$, then 
$$\vspace{-0.5ex} \frac{n+1}{2}=\frac{1}{n}\left( \sum_{y \in P} y  +\sum_{z \in O^{(j-1)}_1} z \right) < \frac{n+1}{2}, \vspace{-0.5ex}$$
which is a contradiction. The case where $\frac{1}{j-1} \sum_{\ell=1}^{j-1} \pi(\ell) \geq \frac{n+1}{2}$ may be handled similarly.
\end{IEEEproof}
We have the following claim.
\begin{claim}\label{cl:edges} Let $\pi \in \cC(n,S_1)$ be constructed according to the previously described running sums $A$. For any integer $1 \leq j \leq n$, 
$$\vspace{-0.5ex} j \cdot \frac{n+1}{2} - (n+1) \leq \sum_{\ell=1}^j \pi(\ell) \leq j \cdot \frac{n+1}{2} + (n+1).\vspace{-0.5ex} $$ \end{claim}
\begin{IEEEproof}
The proof proceeds by induction on $j$. The result holds for $j=1$, and this establishes the base case. 
Suppose next that the result holds for all $j < J$ and consider the case $j=J \leq n$. Clearly,
\begin{align*}\vspace{-0.5ex}
\sum_{\ell=1}^{J-1} \pi(\ell) + \pi(J) \leq \sum_{\ell=1}^J \pi(\ell) \leq \sum_{\ell=1}^{j-1} \pi(\ell)+ \pi(J).\vspace{-0.5ex}
\end{align*}
If 
$\sum_{\ell=1}^{J-1} \pi(\ell) < (J-1) \cdot \frac{n+1}{2},$
then according to Lemma~\ref{ex:d1encode}, we have $\frac{n+1}{2} < \pi(J) \leq n$. Furthermore, using the inductive hypothesis along with $\sum_{\ell=1}^{J-1} \pi(\ell) < (J-1) \cdot \frac{n+1}{2}$ shows that 
$$\vspace{-0.5ex} J \cdot \frac{n+1}{2} - (n+1) \leq  \sum_{\ell=1}^{J} \pi(\ell) \leq J \cdot \frac{n+1}{2} + (n+1), \vspace{-0.5ex}$$
which established the validity of the claim for the case that $\sum_{\ell=1}^{J-1} \pi(\ell) < (J-1) \cdot \frac{n+1}{2}$. 
The case where $\sum_{\ell=1}^{J-1} \pi(\ell) \geq (J-1) \cdot \frac{n+1}{2}$ can be handled similarly.
\end{IEEEproof}

From Claim~\ref{cl:edges}, we can prove that the code $\cC(n,S_1)$ satisfies~(\ref{eq:constraints}).
\begin{lemma}\label{lem:ub} For any $2 \leq b \leq n$ and $j \in [n-b+1]$,
\begin{align*}\vspace{-0.5ex}
b \cdot \frac{n+1}{2}& - 2(n+1) \leq  \pi(j) + \pi(j+1) \\
&+ \ldots + \pi(j+b-1) \leq b \cdot \frac{n+1}{2} + 2(n+1). \vspace{-0.5ex}
\end{align*}
\end{lemma}
\begin{IEEEproof} Clearly,  \vspace{-0.5ex}
\begin{align*}
&\pi(j) + \pi(j+1) + \ldots + \pi(j+b-1) =\\
&\pi(1) + \pi(2) + \ldots + \pi(j+b-1) - \\
&(\pi(1) + \pi(2) + \ldots + \pi(j-1)).\vspace{-0.5ex}
\end{align*}
Using the result of Claim~\ref{cl:edges}, we obtain
\begin{align*} \vspace{-0.5ex}
&\pi(j) + \pi(j+1) + \ldots + \pi(j+b-1) \\
&\leq (j+b-1) \cdot \frac{n+1}{2} + (n+1) - \\
&\left( (j-1) \cdot \frac{n+1}{2} - (n+1) \right) = b \cdot \frac{n+1}{2} + 2(n+1).\vspace{-0.5ex}
\end{align*}
Using the same approach, one may show that 
$$\vspace{-0.5ex}\pi(j) + \pi(j+1) + \ldots + \pi(j+b-1) \geq b \cdot \frac{n+1}{2} -2(n+1),\vspace{-0.5ex}$$
and this completes the proof.
\end{IEEEproof}

These results lead to the following theorem. 
\begin{theorem} 
The capacity of the $D_1$-constraint equals $\cR_1 =1$.
\end{theorem}
\begin{IEEEproof} As a consequence of Lemma~\ref{lem:ub}, we know that the code construction from Example~\ref{ex:d1encode} satisfies the $D_1$-constraint and given that there are $(\frac{n}{2})!$ choices for each of the two permutations $\gamma_1$ and $\gamma_2$, it follows that $\cR_1 = 1$. 
\end{IEEEproof}

Note that a naive implementation of the encoding procedure requires maintaining the sequences $O_1, O_2$ and roughly $O(n^2)$ operations for re-computing the average values of the symbols $n$ times. Clearly, significantly less complex implementations are possible.

One approach would be to divide the input information sequence into two blocks of the same size or sizes that differ by one, and than use the two parts to ``encode'' for the permutations $\gamma_1$ and $\gamma_2$. Here, encoding may refer to generating a permutation at a given position in the lexicographical order of permutations, and efficient, straightforward algorithms for this and more general encodings are known~\cite{Dershowitz75,Erlich73,Knuth74}. This approach would remove the storage requirement for the permutation $\gamma_1$ and $\gamma_2$, and subsequently only require transposing adjacent symbols in the permutations. The procedure, which we next illustrate with an example, may be seen as an extension of Knuth's balancing principle, where complementation used for binary strings is replaced by transpositions in permutations\footnote{There appears to be no natural extension of the notion of complementation in a binary string for permutations, as ``reflecting'' values of a prefix of a permutation around $n+1$ may not result in a permutation.}.

\begin{example} Suppose once more that $n=12$, $P_1=\{1,2,3,4,5,6\},$ $P_2=\{7,8,9,10,11,12 \}$, $\gamma_1=(3,4,1,2,5,6),$ and $\gamma_2 = (6,5,4,3,2,1),$ so that $O_1=(3,4,1,2,5,6)$ and
$O_2=(12,11,10,9,8,7)$.

We form an auxiliary permutation $\pi^{(1)} \in \mathbb{S}_n$ by interleaving $\gamma_1$ and $\gamma_2$, which in the above case leads to
$ \pi^{(1)}  = ({3},\underuparrow{12},\underuparrow{4},11,1,10,2,9,5,8,6,7). $
We maintain two pointers, each requiring $\log(n)$ bits. The initial positions of the pointers are at the location of the second element of $O_1$ and the first element of $O_2$, as we would like to test if transposing these two elements will reduce the running sum. We also initialize the discrepancy to $\Delta(1)=\pi^{(1)}(1)-(n+1)/2 = 3-6.5=-3.5$, and store $\Delta(1)=-3.5$, which requires $O(\log(n))$ bits of overhead.

In the second step of encoding, since $\Delta(1)<0$ and $\pi^{(1)}(2) > \frac{n+1}{2}$, the pointer at $12$ is moved up to the position of the next element in $O_2$ which is $11$ as shown below
$ \pi^{(2)}  = ({3},{12},\underuparrow{4},\underuparrow{11},1,10,2,9,5,8,6,7). $
The updated discrepancy is computed according to $\Delta(2)=\Delta(1)+\pi^{(2)}(2) - 6.5 =2$.

Note that if $\Delta(1) \geq 0$ and $\pi^{(1)}(2) > \frac{n+1}{2}$, then the element pointed at by the second arrow would have been deleted from $\pi^{(1)}$ and reinserted back into the permutation at the position of the first arrow using a single adjacent transposition. At this point, $\pi^{(2)}$ would have had one arrow pointing at the element $1$, the element following $4$ in $O_1$, and another arrow pointing at $12$, the first element in $O_2$.

In the third step, since $\Delta(2) \geq 0$ and $\pi^{(2)}(3) < \frac{n+1}{2}$, no transposition is performed, and we simply move the leftmost pointer to the next element in $O_1$ so that $ \pi^{(3)} =({3},{12},{4},\underuparrow{11},\underuparrow{1},10,2,9,5,8,6,7).$ Note that after three steps of encodings, the positions of three elements in $\pi$ are permanently fixed. We terminate after $n$ elements in $\pi$ have been fixed, in which case we obtain $\pi=(3,12,4,11,1,10, 2, 9, 8, 5, 7,6)$.
 \end{example}


\section{The Capacity of $D_2$-Balanced Codes}

We now consider the case where $D(b,n)$ scales inversely both with $b$ and $\lceil n^{\epsilon} \rceil,$ $\epsilon \in (0,1)$ and where $S_2=\{{2,4,\ldots,2 \cdot s\}}$. The reason behind this choice of problem parameters is that we cannot simultaneously satisfy a stringent discrepancy constraint with $S=[n]$. 
Such a constraint would require that for all $j \in [n]$, one has
$$\vspace{-0.5ex} \frac{n+1}{2} - 8(\lfloor n^{1-\epsilon} + n^{-\epsilon} \rfloor) \leq \pi(j) \leq \frac{n+1}{2} + 8(\lfloor n^{1-\epsilon} + n^{-\epsilon} \rfloor), \vspace{-0.5ex}$$
which is clearly impossible. A similar problem is encountered when $S$ contains two consecutively valued symbols.

Thus, we limit our attention to the case where $S$ contains elements which are multiples of two. The proof follows by noting the $D_2$-constraint requires that for every $i \in [n]$, $\pi(i)$ is close in value to $\pi(i-2)$.

\begin{lemma}\label{lem:ub2} The capacity $\cR_2= 0$ for $s>32(n^{1-\epsilon}+1) + 1$, and $\cR_2 \leq 1-\epsilon$ for $s < 32(n^{1-\epsilon}+1) + 1$. 
\end{lemma}

Throughout the remainder of this section, we write $N = \lceil n^{\epsilon} \rceil$ to ease notational burden and for simplicity assume that $N|n$ and that $N$ is divisible by four. We focus our attention on deriving a lower bound on $\cR_2$ by constructing a balanced code $\cC(n,S_2)$, with  $S_2=\{{2,4,\ldots, 2 \cdot (\frac{n}{N}-1)\}}$ and of rate 
$$\vspace{-0.5ex} \log_{n \to \infty} \frac{ \log | \cC(n,S_2)| }{\log n! } = 1 - \epsilon. \vspace{-0.5ex}$$

We partition the set $[n]$ into $N$ subsets of equal size $\frac{n}{N}$, comprising consecutive integers, subsequently denoted by $P_1$, $P_2, \ldots,$ $P_{N}$\footnote{Clearly, for values of $N$ not satisfying the given divisibility properties, the sets $P_i,\. i=1,\ldots,N$ may have different cardinalities. This small technical detail does not change the validity of the argument nor the claimed result.}.
For each $i \in [ N ]$, we order the elements of $P_i$ arbitrarily and denote the resulting sequence by $O_i$. 
Since there are $\frac{n}{N}!$ ways to arrange each set $P_i$, $| \cC(n,S_2)| \geq \left( \frac{n}{N}! \right)^N$, and hence
$$ \lim_{n \to \infty} \frac{\log \left( \frac{n}{N}! \right)^N}{\log n!} = 1-\epsilon, $$
which implies the lower bound. 

Figure~1 illustrates the encoding process. Similar to what we did before, we incrementally build a permutation $\pi \in \cC(n,S_2)$. Each cell in the figure involves appending two elements to $\pi$, chosen from one of two possible sets. For instance, based on Figure~1, visiting the first cell requires appending either a) one element from $O_2$ and one element from $O_{N}$ or b) appending one element from $O_1$ and one element from $O_{N-1}$ to $\pi$. Note that since each of our sets has size $\frac{n}{N}$, each cell in Figure~1 will be visited $2 \cdot \frac{n}{N}$ times. We next explain this outlined encoding process in more detail.

The  first $2$ symbols of $\pi$ are selected as follows. Set the first element of $\pi$ to be the first element in $O_1$ and then remove this element from $O_1$. Set the second element of $\pi$ to be the first element of $O_{N-1}$ and remove the chosen element from $O_{N-1}$. This selection process is captured by the first cell in Figure~1, indicating that the first element of the permutation is taken from $O_1$ or $O_2$ and the second element is from $O_N$ or $O_{N-1}$. 

We next compute $A(2)=\frac{1}{2} \left( \pi(1) + \pi(2) \right)$ and if $len(O_1)\neq0$ or $len(O_2)\neq0$, we revisit the first cell. If $A(2) < \frac{n+1}{2}$, we append the first element of the set $O_{2}$ followed by the first element of the set $O_{N }$ to $\pi$ and remove these elements from their respective sets. Otherwise, if $A(2) \geq \frac{n+1}{2}$ we append the first element from $O_{1}$ followed by the first element from $O_{N-1}$ to $\pi$ and remove these two elements from their respective sets. We then consider $A(4)=\frac{1}{4} \left( \pi(1) + \pi(2) + \pi(3) + \pi(4) \right)$, and if $len(O_1)\neq0$ or $len(O_2)\neq0$, we revisit the first cell. If $A(4) < \frac{n+1}{2}$, we append the first element from $O_{2}$ followed by the first element from $O_{N}$ to $\pi$ and remove these elements from their respective sets. Otherwise, we append the first element from $O_{1}$ followed by the first element from $O_{N-1}$ to $\pi$ and remove these elements from their respective sets. This process is continued until we have added $4 \cdot \frac{n}{N}$ elements to $\pi$ so that $len(O_1)=len(O_2)=len(O_{N})=len(O_{N-1})=0$. Notice that since two elements are appended at once, we have visited the first cell $2 \cdot \frac{n}{N}$ times.

Next, we again compute the running average of the elements fixed (or appended) to $\pi$ thus far. If the running average is less than $\frac{n+1}{2}$, then we choose the next two elements of $\pi$ to be the first elements of the sets $O_4, O_{N-2}$, which we then remove from their respective sets. Otherwise, the next two elements of $\pi$ are chosen from $O_3,O_{N-3}$, added, and removed from their respective sets. Afterwards, we repeatedly add elements from the sets $O_4, O_{N-2}, O_3, O_{N-3}$ until $len(O_4)=len(O_{N-2})=len(O_3)=len(O_{N-3})=0$. This process is continued until $\pi$ has length $n$.

\begin{figure}[h!]\label{fig:encoding}
  \centering
    \includegraphics[width=0.45\textwidth]{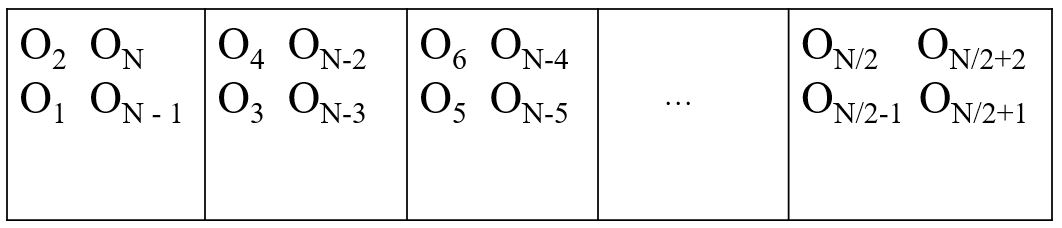}
  \caption{Encoding for a $D_2$-Balanced Code}
\end{figure}
%

We illustrate the procedure with an example.

\begin{example}\label{ex:smaller}  Let $n=32$, $N=8$, and $d=2$, and suppose that 

$O_1=(2,3,4,1),$ $O_2=(8,7,6,5)$, $O_3=(11,10,12,9)$, 

$O_4=(16,13,14,15)$, $O_5=(17,20,19,18)$, 

$O_6=(22,23,21,24)$, $O_7=(25,26,28,27)$, and

$O_8=(32,29,30,31)$.

We set the first two elements of $\pi$ to be the first element from $O_1$ followed by the first element from $O_7$ so that 
$ \pi=(2,25,\ldots) $ 
and $A(2)= \frac{2+25}{2} = 13.5$. Since $A(2) < \frac{n+1}{2}$, we extend $\pi$ to $\pi=(2,25,8,32,\ldots)$. 

 Now, we compute $A(4)=16.75$ which implies that 
$\pi$ should be extended to $\pi=(2,25,8,32,3,26,\ldots)$. 

Next, we find $A(6)=16$ and arrive at 
$\pi=(2,25,8,32,3,26,7,29,\ldots).$
In the next three steps, we compute $A(8)=16.5$, $A(10)=16.4$, $A(12)=16.67$, and $A(14)=16.28$. Based on these values, the permutation $\pi$ is augmented recursively as 
$\pi=(2,25,8,32,3,26,7,29,4,28,\ldots),$
$\pi=(2,25,8,32,3,26,7,29,4,28,6,30,\ldots),$
$\pi=(2,25,8,32,3,26,7,29,4,28,6,30,1,27,\ldots),$
$\pi=(2,25,8,32,3,26,7,29,4,28,6,30,1,27,5,31,\ldots).$ 
At this point, we move onto the second cell in Figure~1. For clarity, we now write $\pi=(\pi^{(1)}, \pi^{(2)}),$ where 
$\pi^{(1)}=(2,25,8,32,3,26,7,29,4,28,6,30,1,27,5,31).$
In the next iteration of the algorithm, we compute $A(16)=16.5$ and augment $\pi^{(2)}$ to $\pi^{(2)}=(11,17)$. In the next four steps we compute $A(18)=16.22$, $A(20)=16.5$, $A(22)=16.36$, $A(24)=16.5$, $A(26)=16.42$, $A(28)=16.5$, and $A(30)=16.3$. The corresponding updates in $\pi^{(2)}$ result in 
$(11,17,16,22,\ldots),$
$(11,17,16,22,10,20,\ldots),$
$(11,17,16,22,10,20,13,23,\ldots),$
$(11,17,16,22,10,20,13,23,12,19,\ldots),$
$(11,17,16,22,10,20,13,23,12,19,14,21,\ldots),$ 
$(11,17,16,22,10,20,13,23,12,19,14,21,9,18,\ldots)$
$(11,17,16,22,10,20,13,23,12,19,14,21,9,18,15,24).$
 \end{example}

Next, we prove that the chosen balancing constraint is satisfied by the previously outlined encoding procedure. The first result in this direction is a generalization of Claim~\ref{cl:edges} from the previous section. It follows directly from the encoding procedure illustrated in Example~\ref{ex:smaller}, and its proof is henceforth omitted.

\begin{claim}\label{cl:edges2} Let $\pi \in \cC(n,S_2),$ where $\cC(n,S_2)$ is constructed according to the described encoding algorithm. 
For any even integer $j\leq n$, it holds that
$$j \cdot \frac{n+1}{2} - \frac{2n}{N} \leq \sum_{\ell=1}^j \pi(\ell) \leq j \cdot \frac{n+1}{2} + \frac{2n}{N}. $$ 
\end{claim}

Suppose now that $j-i+1 \in S_2$. Then, $j-i+1 < 2 \cdot \frac{n}{N}$, and so if $(c-1) \cdot \frac{4n}{N} < i \leq c \cdot \frac{4n}{N},$ then
\begin{align}\label{eq:balancingMUB}
j+1 \leq (c+1) \cdot \frac{4n}{N}.
\end{align}
Thus, if the $i$-th element in a permutation is encoded according to cell index $c_i$ in Figure~1 and $j-i+1 \in S_2$, the $(j+1)$-st element in that permutation is encoded according to cell index $c_i$ or $c_i+1$.  Suppose that $j-i+1 \in S_2$. As a result of the next claim, we know that any element encoded according to cell $c_i$ has a symbol value close to an element encoded according to cell $c_j$.

\begin{claim}\label{cl:edges3} Let $\pi \in \cC(n,S_2)$. Suppose that $i,j$ are such that $j-i+1 \in S_2$. Then,
$$2 | (j-i+1), \;\; (c-1) \cdot \frac{4n}{N} < i < j \leq (c+1) \cdot \frac{4n}{N}$$ 
for some positive integer $c$, and
$$ \pi(i) - \frac{4n}{N} \leq \pi(j+1) \leq \pi(i) + \frac{4n}{N}. $$ \end{claim}


The next lemma establishes that our balancing criteria is satisfied.

\begin{lemma}\label{lem:lower2} For any $i,j \in [n]$ $i<j$, $b=j-i+1 \in S_2$, and $\pi \in \cC(n,S_2)$, we have
\begin{align*}
b \cdot \frac{n+1}{2} - 8 \frac{(n+1)}{N} & \leq  \pi(i) + \pi(i+1) + \ldots + \pi(j) \\
&\leq b \cdot \frac{n+1}{2} + 8 \frac{(n+1)}{N}.
\end{align*}
\end{lemma}
\begin{IEEEproof} Recall from Equation~(\ref{eq:balancingMUB}) and Claim~\ref{cl:edges3} that 
$$(c-1) \cdot \frac{4n}{N} < i < j \leq (c+1) \cdot \frac{4n}{N}$$ 
for some integer $c$. Since $j-i+1 \in S_2$, $j-i+1$ is an even integer. Thus, one of the values $i,j$ is even. Suppose for now that $j$ is even. Then using Claim~\ref{cl:edges2}, we have
\begin{align*}
\pi(i) + \pi(i+1) + \ldots + \pi(j) &= \sum_{\ell=1}^j \pi(\ell) - \sum_{\ell=1}^{i-1} \pi(\ell)\\
&\leq (j-i+1) \cdot \frac{n+1}{2} + \frac{4n}{N}.
\end{align*}
Otherwise, if $j$ is odd, we may write 
\begin{align*}
&\pi(i) + \pi(i+1) + \ldots + \pi(j) = \sum_{\ell=1}^{j+1} \pi(\ell) -  \sum_{\ell=1}^{i} \pi(\ell) + \\
&\pi(i) - \pi(j+1) \leq (j-i+1) \cdot \frac{n+1}{2} +  \frac{4n}{N} + \frac{4n}{N},
\end{align*}
where the inequality follows from Claims~\ref{cl:edges2} and~\ref{cl:edges3}. The inequality in the other direction for the case of $j$ even or $j$ odd can be proved using similar arguments.
\end{IEEEproof}

As a consequence of Lemmas~\ref{lem:ub2} and \ref{lem:lower2}, the following theorem holds.

\begin{theorem} For $\epsilon \in (0,1)$, $R_2 = 1-\epsilon$.
\end{theorem}

\section{The Balanced Two-Neighbor Constraint}
We now turn our attention to a short treatment regarding combined balanced and constrained codes~\cite{FD13}. We will focus on the two-neighbor symmetric constraint coding as defined in~\cite{SY14}. For this purpose, recall that a permutation $\pi \in \mathbb{S}_n$ satisfies the two-neighbor $k$-constraint if for all $i \in \{2,3,\ldots,n-1\}$, either $|\pi(i) - \pi(i-1)| \leq k$ or $|\pi(i) - \pi(i+1)| \leq k$. Let $\cC$ be either a $D_1$- or a $D_2$-balanced permutation code that also satisfies the two-neighbor $k$-constraint. Let
\begin{align}
\cR_i(k) = \lim_{n \to \infty} \sup \frac{ \log | \cC|}{\log n! },
\end{align}
where, similarly to the notation used in the previous sections, $i=1,2$ denotes the capacity of a $D_i$-balanced permutation code that satisfies the two-neighbor $k$-constraint. We consider the case for $D_1$-balanced permutation codes as defined in  (\ref{eq:constraints}) and (\ref{eq:c1}). Recall, for the $D_1$-constraint $S=S_1=[1,n]$. In our derivations, we adopt the same scaling model as in~\cite{SY14}, for which $k=\lceil n^{\epsilon_k} \rceil$ for $\epsilon_k \in (0,1)$. In this case, the capacity is a function of $\epsilon_k$. The discussion of the $D_2$-constrained codes is deferred to an extended version of this paper. 

We pause to note that the $k$-neighbor constraint goes against the balancing methods we outlined so far, as in the latter case we tend to group together symbols with large and small values. 
\begin{theorem} For $\epsilon_k \in (0,1)$, $\cR_1(\lceil n^{\epsilon_k} \rceil) = \frac{1+\epsilon_k}{2}$.\end{theorem}

From~\cite{SY14}, a trivial upper bound on $\cR_1(\lceil n^{\epsilon_k} \rceil)$ is $\frac{1+\epsilon_k}{2}$. Using similar ideas as in Section~\ref{sec:constructd1} and the construction from~\cite{SY14}, one can derive a  matching lower bound. Similarly to what was proposed in~\cite{SY14}, in our setting the $k$-constraint is imposed by selecting two elements at a time from a set with elements that differ by at most $k$. These ideas are illustrated by the next example, while details of the proof are deferred to the full version of the paper. 

\begin{example} Let $n=24$ and $k=4$. We begin by partitioning the set $[24]$ into six sets of size four each, where $P_1=\{1,2,3,4\}$, $P_2=\{5,6,7,8\}$, $P_3=\{9,10,11,12\}$, and $P_4=\{13,14,15,16\}$, $P_5=\{17,18,19,20\}$, $P_6=\{21,22,23,24\}$. We choose an ordering for each of these sets to obtain
\begin{align*}
O_1&=(3,4,1,2), \ \ \  O_2=(8,7,6,5), \ \ \  O_3=(9,10,12,11), \\
O_4&=(13,16,14,15), O_5=(20,19,18,17), O_6=(21,22,23,24).
\end{align*}
We select the first two elements from $O_1$, remove these elements from $O_1$, and arrive at $\pi = (3,4,\ldots)$. 
Next, we compute the running average of $\pi$ as $A(2)=\frac{3 + 4}{2} = 3.5$. Since $3.5 < \frac{n+1}{2} = 12.5$, we choose one of the sets $O_4, O_5, O_6$ to select two additional elements from. Suppose we pick $O_4$ so that $\pi = (3,4,13,16,\ldots$ and so that the symbols $13,16$ are subsequently removed from $O_4$. Then, since the symbol average $A(4)=9 < 12.5$, we choose elements from $O_4,O_5,O_6$. Suppose we pick $O_5$ so that $\pi=(3,4,13,16,20,19,\ldots)$. Then $20,19$ are removed from $O_5$. In this case, $A(6)  \geq 12.5$ and so we pick the next two elements from one of the sets $O_1, O_2, O_3$. Suppose, we choose the set $O_2$ so that $\pi=(3,4,13,16,20,19,8,7$ and so $A(8)=11.25$. Suppose $O_6$ is chosen next and so $\pi=(3,4,13,16,20,19,8,7,21,22$. Suppose we continue the same procedure and choose from the following sets (in order): $O_1, O_5,O_6, O_2, O_3, O_4, O_3$. The result is the permutation $\pi=(3,4,13,16,20,$ $19,8,7,21,22,$ $1,2,18,17,23,24,6$ $,5,9,10,14,15,12,11) \in \mathbb{S}_{24}$.
\end{example}
 
\tiny{}


\begin{thebibliography}{1}

\bibitem{AFG02} 
{R. Anstee, R. Ferguson, and J. Griggs}, ``Permutations with Low Discrepancy Consecutive k-sums,'' \textit{J. of Comb. Theory}, Vol. 100, no. 2, pp. 302-321, Nov. 2002.

\bibitem{SY14}
{S. Buzaglo and E. Yaakobi}, ``Constrained codes for rank modulation,'' \textit{Proc. IEEE ISIT}, Honolulu, HI, Jul. 2014.

\bibitem{Dershowitz75} N. Dershowitz, ``A simplified loop-free algorithm for generating permutations.'' BIT, 15(2):158-164, 1975.

\bibitem{Erlich73} G. Ehrlich, ``Loopless algorithms for generating permutations, combinations, and other combinatorial configurations,'' \emph{Journal of the ACM}, 20(3):500-513, 1973.

\bibitem{JMSB09}
{A. Jiang, R.Mateescu, M. Schwartz, and J.Bruck}, �``Rank modulation for flash memories,'' \textit{IEEE Trans. Inf. Theory}, vol. 55, pp. 2659-2673, Jun. 2009.

\bibitem{Knuth74} D. E. Knuth and J. L. Szwarcfiter. A structured program to generate all topological sorting arrangements. Information Processing Letters, 2:153-157, 1974.

 \bibitem{K86}
{D. E. Knuth}, ``Efficient balanced codes,'' \textit{IEEE Trans. Inf. Theory}, vol. 32, no. 1, pp. 51–53, Jan. 1986.

\bibitem{FD13}
{F. Sala and L. Dolecek}, ``Constrained rank modulation schemes,'' \textit{Proc. IEEE ISIT}, Sevilla, Sept. 2013.

\bibitem{WISS13}
{J.H. Weber, K.A. Immink, P.H. Siegel, T.G. Swart}, ``Perspectives on Balanced Sequences,'' \textit{available at http://arxiv.org/pdf/1301.6484v1.pdf}, Jan. 2013.

\end{thebibliography}
\end{document}